\documentclass[twoside,twocolumn,9pt]{article}
\usepackage{extsizes}
\usepackage[super,sort&compress,comma]{natbib} 
\usepackage[version=3]{mhchem}
\usepackage[left=1.5cm, right=1.5cm, top=1.785cm, bottom=2.0cm]{geometry}
\usepackage{balance}
\usepackage{mathptmx}
\usepackage{sectsty}
\usepackage{graphicx} 
\usepackage{lastpage}
\usepackage[format=plain,justification=justified,singlelinecheck=false,font={stretch=1.125,small,sf},labelfont=bf,labelsep=space]{caption}
\usepackage{float}
\usepackage{fancyhdr}
\usepackage{fnpos}
\usepackage[english]{babel}
\addto{\captionsenglish}{%
  
}
\usepackage{array}
\usepackage{droidsans}
\usepackage{charter}
\usepackage[T1]{fontenc}
\usepackage[usenames,dvipsnames]{xcolor}
\usepackage{setspace}
\usepackage[compact]{titlesec}
\usepackage{hyperref}

\usepackage{bm}
\usepackage{mathtools}
\usepackage{amsmath}
\usepackage{amsfonts}
\usepackage{amssymb}
\usepackage{xfrac}

\usepackage{epstopdf}

\definecolor{cream}{RGB}{222,217,201}

\begin{document}

\pagestyle{fancy}
\thispagestyle{plain}
\fancypagestyle{plain}{
\renewcommand{\headrulewidth}{0pt}
}

\makeFNbottom
\makeatletter
\renewcommand\LARGE{\@setfontsize\LARGE{15pt}{17}}
\renewcommand\Large{\@setfontsize\Large{12pt}{14}}
\renewcommand\large{\@setfontsize\large{10pt}{12}}
\renewcommand\footnotesize{\@setfontsize\footnotesize{7pt}{10}}
\makeatother

\renewcommand{\thefootnote}{\fnsymbol{footnote}}
\renewcommand\footnoterule{\vspace*{1pt}%
\color{cream}\hrule width 3.5in height 0.4pt \color{black}\vspace*{5pt}} 
\setcounter{secnumdepth}{5}

\makeatletter 
\renewcommand\@biblabel[1]{#1}            
\renewcommand\@makefntext[1]%
{\noindent\makebox[0pt][r]{\@thefnmark\,}#1}
\makeatother 
\renewcommand{\figurename}{\small{Fig.}~}
\sectionfont{\sffamily\Large}
\subsectionfont{\normalsize}
\subsubsectionfont{\bf}
\setstretch{1.125} 
\setlength{\skip\footins}{0.8cm}
\setlength{\footnotesep}{0.25cm}
\setlength{\jot}{10pt}
\titlespacing*{\section}{0pt}{4pt}{4pt}
\titlespacing*{\subsection}{0pt}{15pt}{1pt}

\fancyfoot{}
\fancyfoot[LO,RE]{\vspace{-7.1pt}\includegraphics[height=9pt]{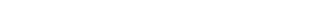}}
\fancyfoot[CO]{\vspace{-7.1pt}\hspace{13.2cm}\includegraphics{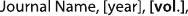}}
\fancyfoot[CE]{\vspace{-7.2pt}\hspace{-14.2cm}\includegraphics{head_foot/RF}}
\fancyfoot[RO]{\footnotesize{\sffamily{1--\pageref{LastPage} ~\textbar  \hspace{2pt}\thepage}}}
\fancyfoot[LE]{\footnotesize{\sffamily{\thepage~\textbar\hspace{3.45cm} 1--\pageref{LastPage}}}}
\fancyhead{}
\renewcommand{\headrulewidth}{0pt} 
\renewcommand{\footrulewidth}{0pt}
\setlength{\arrayrulewidth}{1pt}
\setlength{\columnsep}{6.5mm}
\setlength\bibsep{1pt}

\makeatletter 
\newlength{\figrulesep} 
\setlength{\figrulesep}{0.5\textfloatsep} 

\newcommand{\topfigrule}{\vspace*{-1pt}%
\noindent{\color{cream}\rule[-\figrulesep]{\columnwidth}{1.5pt}} }

\newcommand{\botfigrule}{\vspace*{-2pt}%
\noindent{\color{cream}\rule[\figrulesep]{\columnwidth}{1.5pt}} }

\newcommand{\dblfigrule}{\vspace*{-1pt}%
\noindent{\color{cream}\rule[-\figrulesep]{\textwidth}{1.5pt}} }

\makeatother

\twocolumn[
\begin{@twocolumnfalse}
{
\hfill\raisebox{0pt}[0pt][0pt]
{\includegraphics[height=55pt]{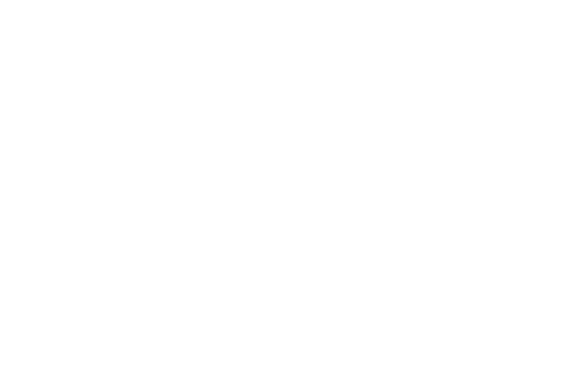}}\\[1ex]
\includegraphics[width=18.5cm]{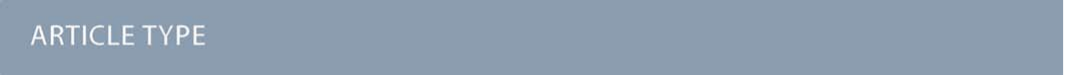}}\par
\vspace{1em}
\sffamily
\begin{tabular}{m{4.5cm} p{13.5cm} }

\includegraphics{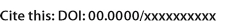} & \noindent\LARGE{\textbf{Optothermally Induced Active and Chiral Motion of the Colloidal Structures$^\dag$}} \\
\vspace{0.3cm} & \vspace{0.3cm} \\

 & \noindent\large{Rahul Chand,\textit{$^{a, \ddag}$}, Ashutosh Shukla,\textit{$^a$} Sneha Boby \textit{$^a$}} and G V Pavan Kumar,$^{a, \ast}$ \\

\includegraphics{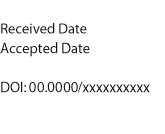} & \noindent\normalsize{Artificial soft matter systems have appeared as important tools to harness mechanical motion for microscale manipulation. Typically, this motion is driven either by the external fields or by mutual interaction between the colloids. In the latter scenario, dynamics arise from non-reciprocal interaction among colloids within a chemical environment. In contrast, we eliminate the need for a chemical environment by utilizing a large area of optical illumination to generate thermal fields. The resulting optothermal interactions introduce non-reciprocity to the system, enabling active motion of the colloidal structure. Our approach involves two types of colloids: passive and thermally active. The thermally active colloids contain absorbing elements that capture energy from the incident optical beam, creating localized thermal fields around them. In a suspension of these colloids, the thermal gradients generated drive nearby particles through attractive thermo-osmotic forces. We investigate the resulting dynamics, which lead to various swimming modes, including active propulsion and chiral motion. We have also experimentally validated certain simulated results. By exploring the interplay between optical forces, thermal effects, and particle interactions, we aim to gain insights into controlling colloidal behavior in non-equilibrium systems. This research has significant implications for directed self-assembly, microfluidic manipulation, and the study of active matter.} 

\end{tabular}

 \end{@twocolumnfalse} \vspace{0.6cm}

  ]

\renewcommand*\rmdefault{bch}\normalfont\upshape
\rmfamily
\section*{}
\vspace{-1cm}

\footnotetext{\textit{$^a$~ Department of Physics, Indian Institute of Science Education and Research (IISER) Pune, Pune, 411008, India}}

\footnotetext{\textit{$^{\ddag}$~rahul.chand@students.iiserpune.ac.in}}
\footnotetext{\textit{$^{\ast}$~pavan@iiserpune.ac.in}}

\footnotetext{\dag~Electronic supplementary information (ESI) available. \url{https://drive.google.com/drive/folders/1VT2lkM6BqkX1lu6vUXnMUAYOkZOORJ5t?usp=sharing}}



\section*{Introduction}
Out-of-equilibrium soft matter systems display a range of complex behaviors including active propulsion \cite{fodor2016far, speck2016stochastic, takatori2015towards, bowick2022symmetry, needleman2017active, van2018grand, hallatschek2023proliferating, de2015introduction, fodor2018statistical, ramaswamy2010mechanics, sanchez2012spontaneous, weber2011active}, self-organization \cite{tan2022odd, singh2017non, araujo2023steering}, and other emergent dynamics \cite{budrene1995dynamics, yang2024shaping, curatolo2020cooperative, steager2008dynamics, theeyancheri2024dynamic}. These phenomena arise as a result of the mutual interaction between the individual element and external stimuli \cite{speck2020collective, takatori2014swim}. Artificial synthetic colloidal systems provide a controlled test bed for studying this behavior \cite{wang2020practical}. Various external fields such as optical \cite{peng2020opto, raj2025direct, nedev2015optically, tkachenko2023evanescent, simoncelli2016combined, buttinoni2012active, haldar2012self, kumar2024inhomogeneous, roy2013controlled}, chemical \cite{buttinoni2012active, simoncelli2016combined, kumar2022chemically}, electrical \cite{zhang2012directed, erez2022electrical, chen2018electrically, demirors2018active,komazaki2015electrically} and magnetic fields \cite{tierno2008controlled, tierno2008magnetically, demirors2018active, komazaki2015electrically} are commonly used as manipulation techniques. Most of the studies are based on inherently asymmetric systems where the structural or compositional asymmetry induces force imbalance leading to the active motion of the structure \cite{moyses2016trochoidal, jiang2010active, wang2018cu, stanton2016biohybrid, simmchen2016topographical, kummel2013circular, raj2025direct, mondal2015generation, theeyancheri2020translational}.

On the contrary, researchers have proposed the use of symmetrical colloids with different chemical reactivity to enable active motion without inherent asymmetry \cite{yu2018chemical, schmidt2019light, grauer2021active, varma2018clustering, codina2022breaking, kumar2020pitch}. In these cases, the dynamics of the colloidal structures are powered by the non-reciprocal interactions. However, these methods require chemical environments such as hydrogen peroxide \cite{yu2018chemical} or water-lutidine mixtures \cite{ schmidt2019light, grauer2021active}. Such requirements restrict the usability of these platforms.

\begin{figure*}[ht]
    \centering
    \includegraphics[width = 525 pt]{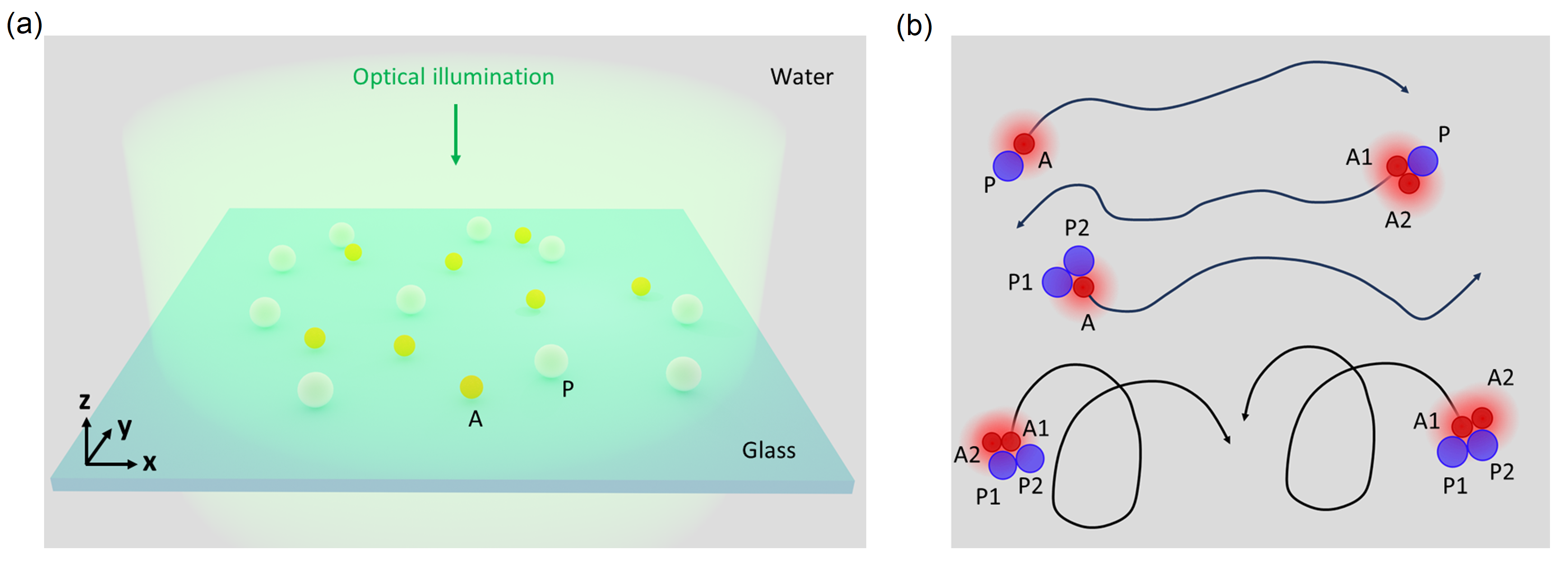}
    \caption{Schematic of the dynamics. (a) When an aqueous colloidal suspension of thermally active (A) and passive (P) colloids is illuminated by a large optical field, the thermally active colloids absorb the laser energy and heat up. This results in a temperature difference between the colloids. (b) The temperature difference induces thermo-osmotic interactions between the colloids. As a result, different colloidal structures, such as dimers, trimers, and quadromers, are formed, and they exhibit various swimming modes.}
    \label{Figure_1}
    
\end{figure*}

Regarding this, some recent studies show that in an aqueous medium, colloidal particles can be dragged to the heat center by the thermo-osmotic slip flow produced by it \cite{bregulla2016thermo, franzl2022hydrodynamic, quinn2023thermofluidic}. One can utilize this process to harness non-reciprocal attractive interactions without any chemical environment. In systems with absorbing and non-absorbing colloids, these interactions induce imbalance and drive the motion \cite{chand2023emergence}. Recently, we have reported the directional motion obtained through such thermo-osmotic interactions in various optical confinement (such as Gaussian, line, and ring beam) \cite{chand2023emergence, chand2024optothermal}. The interaction between the absorbing and non-absorbing colloids fuels the dynamics. However, the induction of similar dynamics in an unconfined environment is yet to be explored in detail. These may have broad applications in areas such as directed self-assembly \cite{chung2008guided}, microfluidic manipulation \cite{jae2018microfluidic, hettiarachchi2023recent, rhee2008microfluidic, schneider2011algorithm}, and the exploration of non-equilibrium processes in active matter \cite{davis2024active}.

Motivated by this, in this article, we use numerical simulations to explore the dynamics of colloidal structures made of absorbing thermally active (A) and non-absorbing passive (P) colloids under broad-area optical illumination (see Fig. \ref{Figure_1}(a)). Such illumination generates temperature gradients without any confinement potential. We examine how these gradients affect particle motion and interactions, leading to different swimming modes (see Fig. \ref{Figure_1}(b)). Our experimental results support these findings. This study advances the understanding of the control of colloidal dynamics through optical and thermal manipulation in non-equilibrium systems \cite{bechinger2016active}.

\begin{figure*}[ht!]
    \centering
    \includegraphics[width = 525 pt]{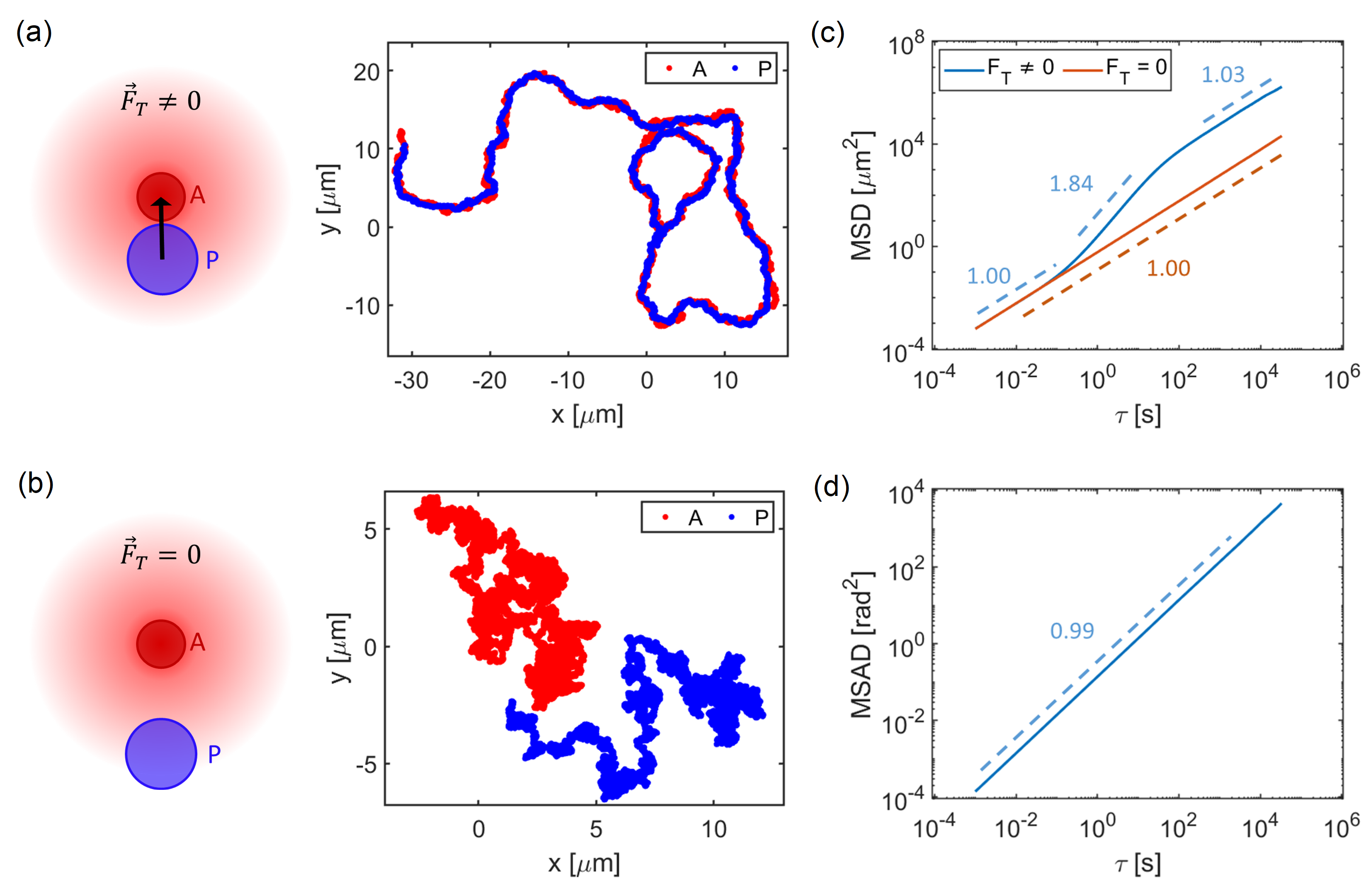}
    \caption{Dynamics of colloidal dimer under large area optical illumination. The attractive thermo-osmotic interaction between one thermally active (A) and a passive (P) colloid leads to the formation of an active dimer (AD) structure. (a) The trajectory of such an AD is shown. The trajectory of the colloids without any thermo-osmotic interaction is shown in (b). The corresponding mean squared displacement (MSD) for (a) and (b) is shown in (c). The MSD undergoes diffusive to ballistic and, again, diffusive transition with increasing timescales. In the absence of interaction, the colloids undergo random diffusion, and the MSD scales linearly, indicating the role of the thermo-osmotic interaction for the formation of AD and the resulting active propulsion. The orientation vector for the active dimer structure is illustrated by the black arrow in (a). (d) The mean squared angular displacement (MSAD) the orientation vector indicates that although the AD undergoes active motion, its orientation shows a diffusive trend.}
    \label{Figure_2}
    
\end{figure*}

\section*{Simulation model}

\begin{figure*}[ht]
    \centering
    \includegraphics[width = 525 pt]{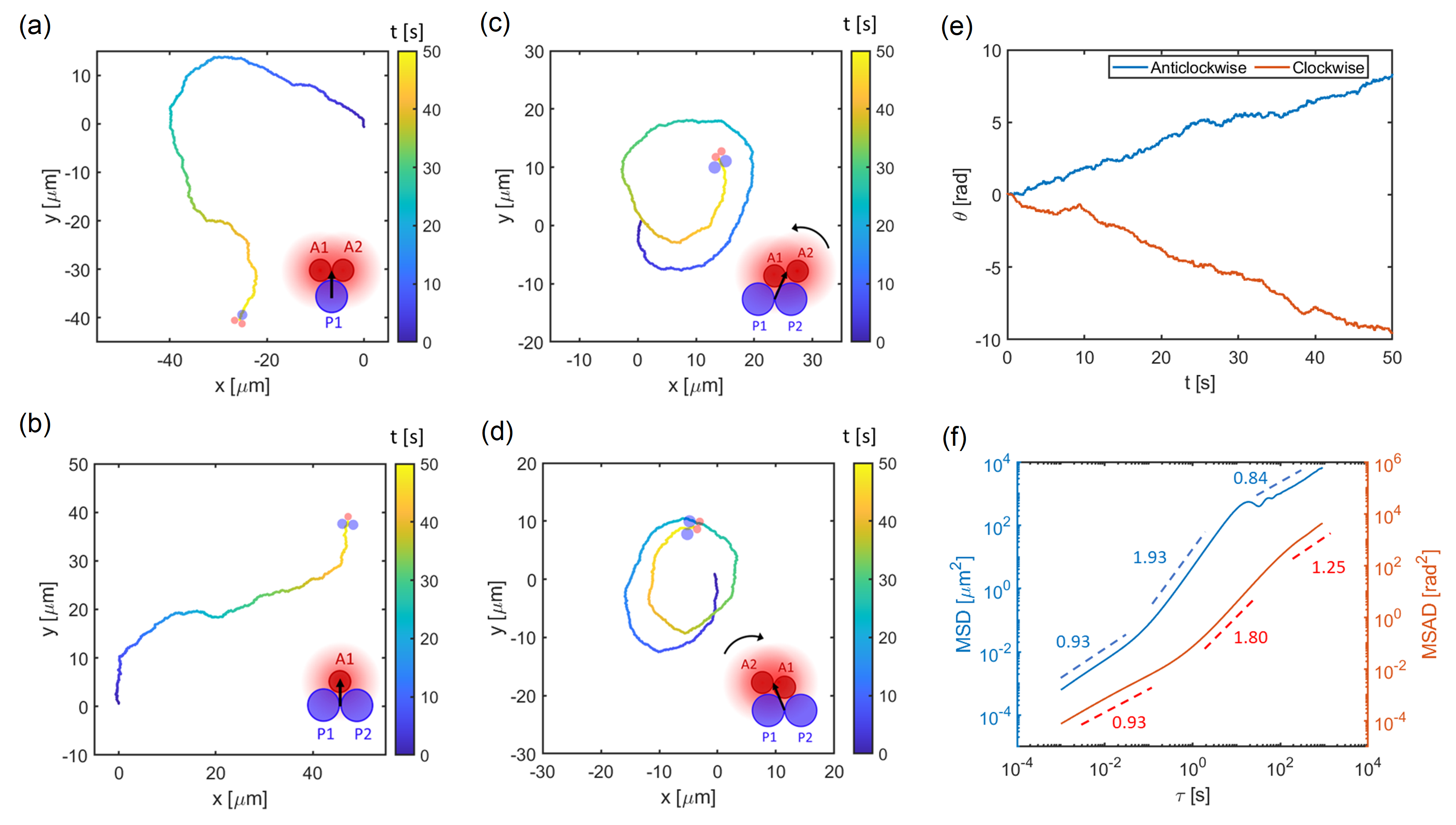}
    \caption{Dynamics of active trimer and quadromer structures and the emergence of chirality. Two possible trimer configurations are: (a) containing one passive (P1) with two thermally active colloids (A1, A2) and (b) two passive (P1, P2) with one thermally active colloid (A1). The trajectories of both these trimer configurations exhibit only active propulsion without any preferred chirality or handedness. In contrast, the active quadromer structures composed of two passive (P1, P2) and two thermally active colloids (A1, A2) exhibit (c) anticlockwise and (d) clockwise handedness based on their respective arrangement of the constituent colloids. (e) The corresponding time evolution of the orientation vector for active quadromers, shown in (c) and (d). (f) The mean squared displacement (MSD) and the mean squared angular displacement (MSAD) indicate the activity signature in both linear and angular coordinate.}
    \label{Figure_3}
\end{figure*}

In our study, we have considered thermally active and passive colloids to investigate the dynamics arising from the temperature difference between the colloids. The thermally active colloids are absorbing, and the passive colloids are non-absorbing in nature. The resultant dynamics of the system are studied by solving the coupled Langevin equation, which describes the influence of various forces and thermal noise on the motion of colloids. The Langevin equation governing the dynamics can be expressed as:

\[
\gamma_i \vec{v}_i = \vec{F}_i^{ext} + \vec{\eta}_i
\]

Where, \( \vec{v}_i \) represents the instantaneous velocity of colloid \( i \), \( \gamma_i \) is the viscous drag coefficient, \( \vec{F}_i^{ext} \) denotes the external force acting on colloid \( i \), and \( \vec{\eta}_i \) accounts for the thermal noise-induced force. The external force \( \vec{F}_i^{ext} \) encompasses contributions from both the external optical field \( \vec{F}_i^o \) and the mutual interaction forces \( \vec{F}_{ij}^{int} \) arising from other colloids \( j \). Additionally, the body forces between colloids are modeled using Lennard-Jones interaction forces, denoted as \( (\text{LJ})_{ij} \). Therefore, we can express the external force as:

\[
\vec{F}_i^{ext} = \vec{F}_i^o + \sum_{j\neq i} \vec{\text{LJ}}_{ij} + \sum_{j\neq i} \vec{F}_{ij}^{int}
\]

Our primary interest lies in the colloidal dynamics resulting from mutual optothermal interactions under broad optical illumination. In this context, we can simplify our analysis by neglecting the optical force component, allowing us to rewrite the external force term as:

\[
\vec{F}_i^{ext} = \sum_{j\neq i} \vec{\text{LJ}}_{ij} + \sum_{j\neq i} \vec{F}_{ij}^{int}
\]

When the colloids are subject to thermal fields, the significant interaction documented in the literature is through thermo-osmotic interactions \cite{bregulla2016thermo, franzl2022hydrodynamic, quinn2023thermofluidic}. These interactions occur when a local thermal field induces a thermo-osmotic slip flow \( \vec{v}_s \) in the vicinity of the colloids, potentially dragging nearby colloids toward the heated colloids. The interaction force can be defined as:

\[
\vec{F}_{ij}^{int} = \vec{F}_{ij}^{TO} = \gamma_i \vec{v}_{s,j}
\]

Here, \( \vec{F}_{ij}^{TO} \) indicates the thermo-osmotic interaction force on colloid \( i \) due to the thermo-osmotic flow \( (\vec{v}_{s,j}) \) produced by colloid \( j \). As the passive colloids are non-absorbing in nature, they cannot induce thermal fields and the consequent thermo-osmotic flow, resulting in: 

\[
\vec{v}_{s,j} = 0  \quad  \Rightarrow  \quad  \vec{F}_{ij}^{int} = \vec{F}_{ij}^{TO} = \gamma_i \vec{v}_{s,j} = 0
\]

Taking all these considerations into account, we can formulate the coupled Langevin equation as follows:

\[
\gamma_i \vec{v}_i = \vec{\eta}_i + \sum_{j\neq i} \vec{\text{LJ}}_{ij} + \sum_{j\neq i} F_{ij}^{int}
\]

By numerically solving this equation, we can explore the dynamics that emerge in a colloidal suspension composed of both thermally active (A) and passive (P) colloids under broad optical fields.

\section*{Results and discussion}
\subsection*{Active motion of dimer structure}

\begin{figure*}[ht]
    \centering
    \includegraphics[width = 400 pt]{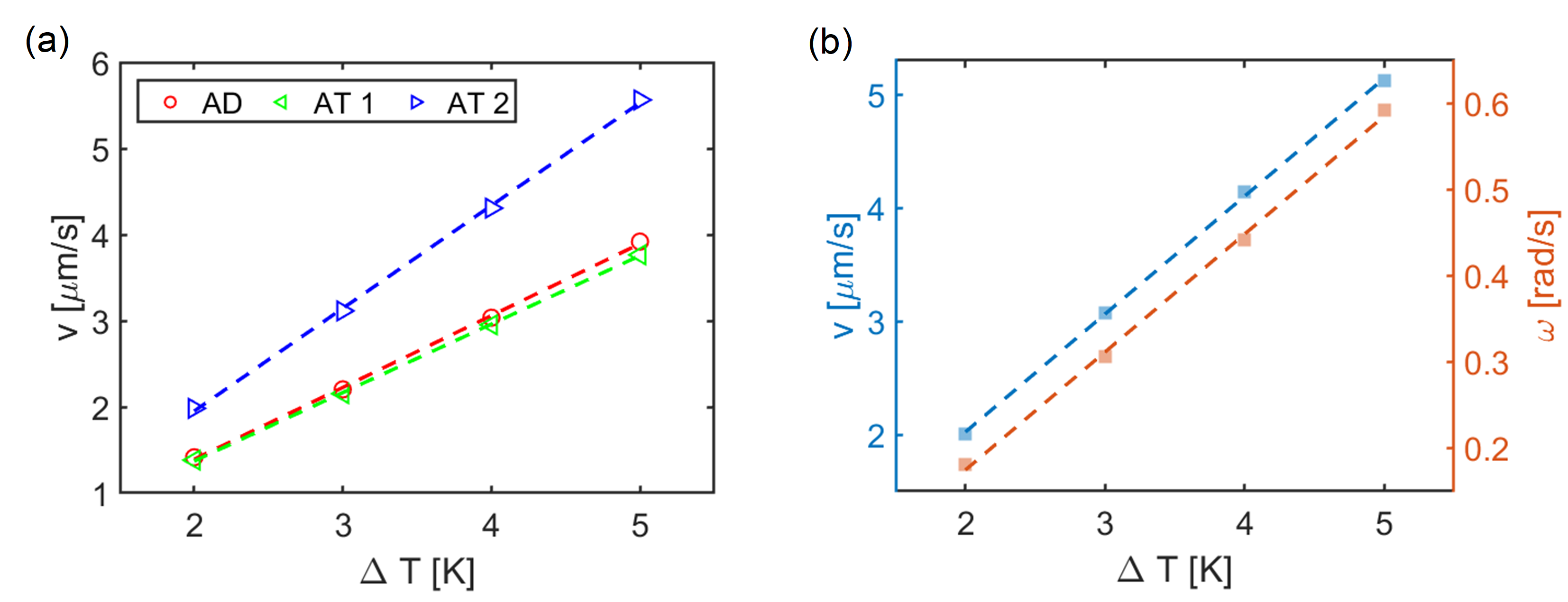}
    \caption{Influence of magnitude of temperature difference on the dynamics of the active structures. An incremental trend of the linear velocities (v) of the active dimer (AD) and both active trimer structures (AT1 and AT2) are shown in (a). The corresponding incremental trend of the linear (v) and angular velocities ($\omega$) for the active quadromer structures are shown in (b). The dashed lines indicated the linear fit.}
    \label{Figure_4}
    
\end{figure*}

Under broad optical illumination, thermally active colloids effectively harness laser energy and dissipate heat. This process generates a localized temperature distribution around the thermally active colloids. This thermal field results in thermo-osmotic slip fluid flow, which acts as an environmental cue to manipulate the motion of nearby colloidal particles. Specifically, when the thermally active colloids interact with their passive counterpart, these thermo-osmotic interactions become non-reciprocal, forming various bound structures. The simplest possibility is a dimer structure composed of one thermally active (A) and passive (P) colloid. Due to the non-reciprocity of the interactions, these dimer structures exhibit propulsion behavior as illustrated in the trajectory depicted in Fig. \ref{Figure_2}(a). In contrast, when the attractive thermo-osmotic interaction force is ignored, the constituent colloids revert to their independent diffusive motion. This is evident from the trajectory shown in Fig. \ref{Figure_2}(b), where each colloid follows its own diffusive path and does not exhibit any correlated motion (see SI video 1). We have calculated the mean-squared displacement (MSD) to quantify these dynamics, as presented in Fig. \ref{Figure_2}(c). The MSD analysis reveals distinct behaviors at varying timescales: at very short timescales, the dimer exhibits a diffusive linear trend; at moderate timescales, it transitions to ballistic motion; and at very large timescales, it again displays diffusive characteristics. This behavior is emblematic of typical active systems \cite{volpe2014simulation}. However, when the attractive thermo-osmotic interaction between the colloids is ignored, the MSD remains linear across all timescales, indicating the absence of activating factors. This observation suggests the vital role of slip flow-induced non-reciprocal attractive interactions in sustaining the activity. The dimer structure propels without any chemical environment. Furthermore, we can associate a specific orientation direction with this active dimer structure, represented by a black arrow connecting the centers of the two colloids in Fig. \ref{Figure_2}(a). To investigate how the orientation of this active dimer structure evolves, we estimated the mean squared angular displacement (MSAD) of this directional vector, as shown in Fig. \ref{Figure_2}(d). Slope 1 suggests that the orientation of the active dimer lacks any inherent rotational bias.

\begin{figure*}[ht]
    \centering
    \includegraphics[width = 525 pt]{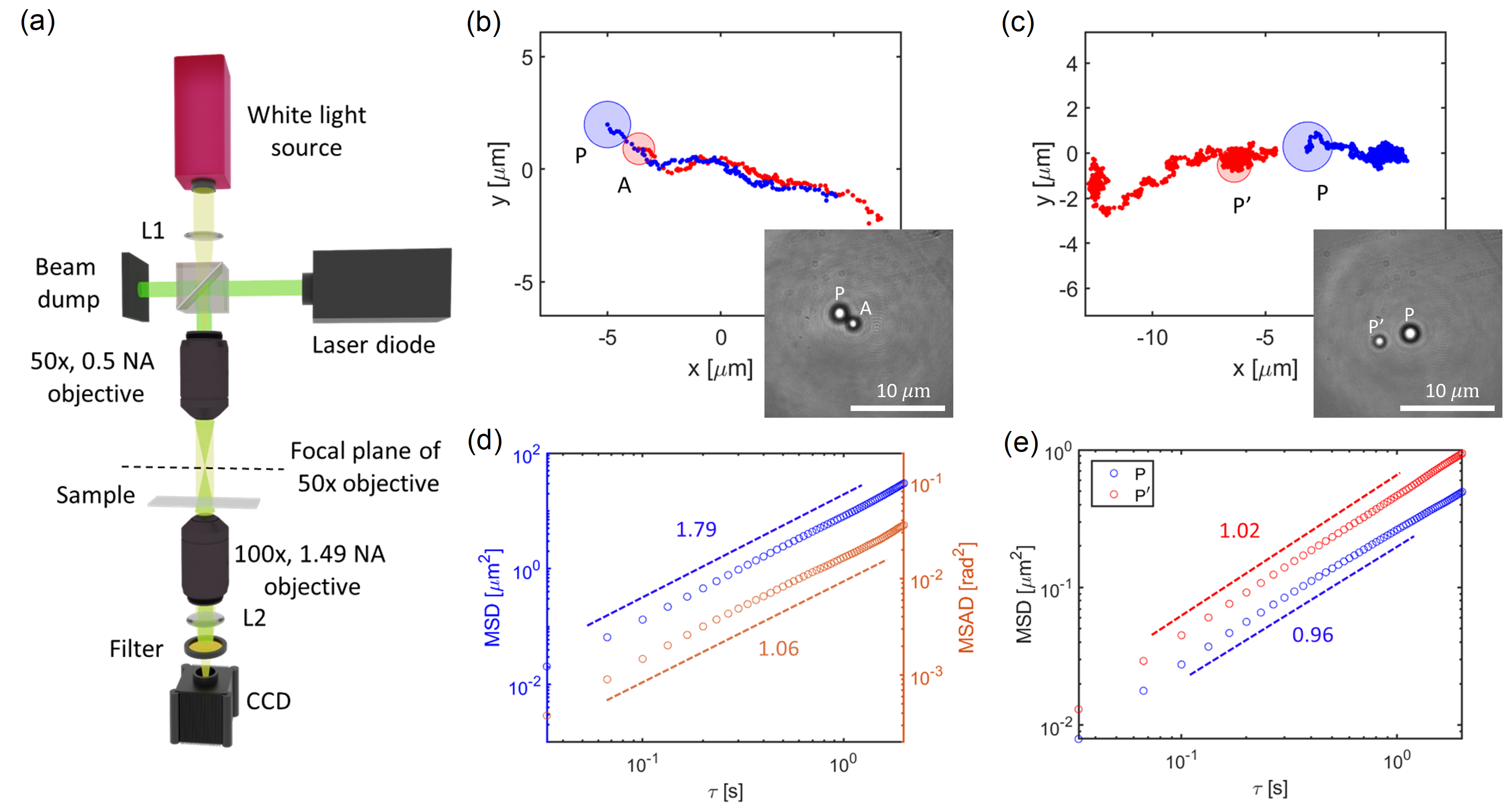}
    \caption{Experimental validation of the active propulsion. (a) Experimental setup used for the experiments. (b) The trajectory of an active dimer structure composed of one passive (P) and one thermally active colloid (A). (c) The trajectory of a dimer structure when the thermally active colloid (A) is replaced by a passive colloid (P') of the same composition and size. The corresponding bright field optical images are shown in the inset. (d) The average mean squared displacement (MSD) of the active dimer structures of dimer structure clearly shows that the slope of the curve is 1.79, which suggests that the motion is ballistic. The 1.06 slope of the mean squared angular displacement (MSAD) indicates that the orientation vector is diffusive. (e) The corresponding slope of the mean squared displacement for the colloids in (c) is close to 1, indicating the colloids undergoing diffusive motion in the absence of the thermo-osmotic interaction.}
    \label{Figure_5}
    
\end{figure*}

\subsection*{Inducing rotational bias}
To gain deeper insights into the dynamics that emerge from evolving colloidal structures due to temperature differences, we simulated the dynamics of trimer and quadromer configurations. There are two distinct configurations, AT1 and AT2. (i) AT1 contains one thermally active colloid (A1) paired with two passive colloids (P1 and P2) and (ii)  AT2 contains two thermally active colloids (A1 and A2) alongside one passive colloid (P1). In both these trimer configurations, the non-reciprocal thermo-osmotic interaction can only lead to an imbalance of effective force along the structure's symmetry line. As a result, they both exhibit propulsion behavior similar to the active dimer structure (see SI video 2). Fig. \ref{Figure_3}(a) and \ref{Figure_3}(b) show the corresponding time-evolving trajectories of these two trimer configurations. We can clearly observe from the trajectory that although these trimer structures undergo active propulsion, they also do not show any preferred angular motion (see supplementary information 1). 

In contrast, if we consider a quadromer structure composed of two passive and thermally active colloids (as shown in Fig. \ref{Figure_3}(c)), due to the respective arrangement of the colloids, the passive colloid P1 is relatively more distant than the passive colloid P2 from the thermally active colloid A2. This leads to different magnitudes of attractive thermo-osmotic force acting on the passive colloids. The thermo-osmotic force on P2 is more than P1. As a result, P2 tends to move faster than P1. But since the quadromer moves together, the trajectory tends to move toward the colloid P1, resulting in an anticlockwise chiral motion, as depicted in Fig. \ref{Figure_3}(c). Similarly, these active quadromer structures can exhibit clockwise motion, as shown in trajectory Fig. \ref{Figure_3}(d), by altering the relative location of the thermally active colloids (see SI video 3). We can define an orientation vector of the colloidal quadromer along the line joining from the center of mass of the passive colloids P1 and P2 to the center of mass of the thermally active colloids A1 and A2. The time evolution of the orientation vector clearly indicates the persistent angular motion in the case of both chiral quadromers, as shown in Fig. \ref{Figure_3}(e). These chiral quadromers can reconfigure the constituent colloids due to the stochastic Brownian noise and can switch from one kind of handedness to another (see SI video 4). The calculated mean squared displacement and the mean squared angular distribution vector for such a chiral quadromer structure are shown in Fig. \ref{Figure_3}(f).  In contrast to the active dimer and trimer structure, in this case, we observed a transition from linear to quadratic and, again, linear in both MSD and MSAD. This suggests that although the structures exhibit preferred handedness at very long timescales, both the handedness of the colloids are equally probable and will effectively show diffusive behavior in both length and angular scales. Similar chiral motion is also observed in natural active systems \cite{vincenti2019magnetotactic, lauga2006swimming, erglis2007dynamics, dominick2018rotating}. Similar dynamics have been observed and studied in artificial colloidal structures with inherent asymmetric structures or with chemical environments \cite{jiang2010active, schmidt2019light, liebchen2022chiral, schmidt2018microscopic, peng2020opto, caprini2023chiral, muzzeddu2022active, donato2016light}. But here, we show the dynamics in symmetric colloids with a non-chemical environment purely due to the optothermal interaction.

\subsection*{Control of the dynamics by the magnitude of temperature difference}
In the previous sections, we have shown how 
the temperature difference between the colloids and the resulting thermo-osmotic interaction leads to active propulsion and chiral motion in different colloidal structures. To understand the change in dynamics with the magnitude of the temperature difference, we simulated the dynamics with four different mismatch values. These temperature differences can be altered in real geometry by changing the intensity of the incident in a broad optical field. Our results show that the active velocities of the dimer, trimer, and quadromer structures exhibit a clear incremental trend with increasing temperature difference, as indicated in Fig. \ref{Figure_4} (a) to \ref{Figure_4}(b). Additionally, the angular velocity of the quadromer structure displays a similar trend, indicated in \ref{Figure_4}(b). The dashed lines in these figures represent corresponding linear fits. This suggests both active propulsion and angular velocities scale linearly with temperature difference values owing to the variation in the intensity of the incident optical field.

\subsection*{Experimental validation}
We conducted a series of experiments to validate our simulated results. We used melamine formaldehyde (MF) colloids with a diameter of 2 {\textmu}m  as passive (P) colloids and nano-particulate iron-oxide infused polystyrene (PS) colloids with a diameter of 1.3 \textmu m as thermally active (A) colloids. Broad optical illumination is essential to replicate the dynamics observed in our simulations. However, our laser diode produced relatively limited optical output. So, instead of a uniform broad illumination, we employed a defocused optical illumination with intensity 64 $\mu \textrm{W}/\mu \textrm{m}^2$ to induce a temperature difference between the thermally active and passive colloids. Our experimental design utilized a dual-channel optical setup, as shown in Fig. \ref{Figure_5}(a). A 532 nm continuous-wave Gaussian laser beam served as the light source, which was coupled to the sample plane using a 50x, 0.5 NA objective lens. Additionally, a white light source was integrated into the setup via a beam splitter to facilitate visualization of the dynamics. The sample plane was positioned away from the objective lens's focal plane (F, dashed line in Fig. \ref{Figure_5}(a)), resulting in a defocused larger illuminated area compared to while positioning it at the focus. The defocused beam heats the thermally active colloids, setting up a temperature gradient between the thermally active and passive colloids. These temperature gradients enable the resulting thermo-osmotic interactions. To capture the dynamics of the colloids, we collected signals from the sample plane using a 100x, 1.49 NA oil immersion objective, which were then projected onto the sensor of a CCD camera. The trajectories of the colloidal particles were extracted from recorded video files using the TrackMate \cite{tinevez2017trackmate} extension of Fiji software \cite{schindelin2012fiji}. The defocused optical field harnessed the temperature difference between the colloids and facilitated thermo-osmotic interactions that formed dimer structures exhibiting active dynamics (SI video 5). A representative trajectory of such a dimer structure is illustrated in Fig. \ref{Figure_5}(b), which corroborates our simulated results of the coupled active motion of dimer structures formed by one passive (P) and one thermally active colloid (A). If we replace the thermally active colloid with a passive colloid (P') of the same composition and size, then no tightly bound dimer structure that exhibits active motion is formed (see SI video 6). Thus, evidence is provided to support our claim. This is because the passive polystyrene colloids (P’) do not possess any absorbing agent like the nanoparticulate iron oxide in the thermally active colloids. They lack the ability to produce the thermal field. Thus, the attractive thermo-osmotic interaction is absent in this case, and as a result, the colloids do not form tightly bound dimer structures and undergo diffusive motion individually, as indicated by the trajectory in Fig. \ref{Figure_5}(c). The corresponding logarithmic plot of the experimental mean squared displacement (MSD) of the active dimer and the passive dimer are shown in Fig. \ref{Figure_5}(d) and (e), respectively. The experimental slope of the MSD of the active dimer is 1.79, which is close to 2. In contrast, for the passive dimer, the slope is close to 1 for the individual particles. This highlights the active motion of the dimer structure, which is facilitated by the non-reciprocal interaction between the constituent colloids. The experimental propulsion velocity of the active dimer structures can be obtained by fitting the experimental MSD with the equation $MSD(\tau) = 4D\tau + v^2\tau^2$, where D and v are the diffusivity and propulsion velocity of the structures \cite{jiang2010active}. We obtain the propulsion velocity as v = 2.60 {\textmu}m/s. We have also noticed that this active velocity increases with increasing intensity of the optical illumination (see supplementary information 2). To validate that the active dimer structures do not show any preferred angular motion, we calculated the mean squared angular displacement (MSAD) of the orientation vector (as shown in (Fig. \ref{Figure_5}(d)) and found that the slope is close to 1, suggesting the diffusive nature. Furthermore, we have also observed similar active motion signatures in trimer colloidal structures, as predicted by our simulated results (see SI video 7). However, due to the limited optical illumination over a limited area of the sample plane, we were not able to observe the active motion displayed by active quadromers. However, we anticipate that similar dynamics can also be observed experimentally using a proper broad area of optical illumination with sufficient optical intensity.

\section*{Conclusions}
Thus, in this study, we have explored the intricate behaviors of colloidal structures composed of thermally active and passive colloids by solving coupled Langevin equations. Our investigation revealed various emerging swimming modes, including active propulsion and chiral handedness, which are primarily driven by temperature differences between the constituent colloids resulting from an attractive nonreciprocal interaction between the colloids as we found that these emerging dynamics are absent when the non-reciprocal attractive interactions are ignored. 

In contrast to the available literature, we did not rely on any chemical environment, such as hydrogen peroxide or 2,6-Lutidine, to harness the nonreciprocal interaction \cite{yu2018chemical, schmidt2019light, grauer2021active, varma2018clustering, codina2022breaking}. In our study, the non-reciprocal interaction between the colloids is employed by the thermo-osmotic slip fluid flow due to the temperature difference of the colloids. These interactions result in an inherent force imbalance within the structures, leading to both propulsion and chiral motion influenced by the constituent colloidal particles. Additionally, we observed that the dynamics of the colloidal structures can be enhanced or diminished by tuning the magnitude of the temperature mismatch, achievable through uniform optical illumination across a large area and by controlling the intensity of the illumination. We have experimentally validated the active dynamics of colloidal structures where a defocused laser illumination maintains the temperature difference between colloids.

In addition, since the dynamics emerge due to thermo-osmotic flow produced by the heated colloid in aqueous solution, this platform is potentially useful for cargo transport. This study may also contribute valuable insights into the dynamics of colloidal structure, paving the way for potential applications in micro-engineering \cite{xupower, galajda2001complex, rubinsztein2002light} and microfluidic devices \cite{jae2018microfluidic, hettiarachchi2023recent, rhee2008microfluidic, schneider2011algorithm}.

\section*{Author contributions}
R.C., A.S., and G.V.P.K designed the study. R.C. and S.B. performed the experiments and analysis. A.S. has helped in the Brownian dynamics simulations. All the authors wrote the manuscript draft together.

\section*{Conflicts of interest}
There are no conflicts to declare.


\section*{Acknowledgements}
 R.C. thanks Diptabrata Paul and Sumant Pandey for the fruitful discussion. This work was partially funded by AOARD (grant number FA2386-23-1-4054) and the Swarnajayanti
fellowship grant (DST/SJF/PSA-02/2017-18) to G.V.P.K.



\balance


\bibliography{softmatter.bib} 
\bibliographystyle{rsc} 

\end{document}